\documentclass[twocolumn,showpacs,aps,prl,superscriptaddress]{revtex4}

\usepackage{graphicx}
\usepackage{dcolumn}
\usepackage{amsmath}
\usepackage{epsfig}

\input pubboard/babarsym

\def\kaons  {\ensuremath{K^{(*)}}\xspace}

\def\pion  {\ensuremath{\pi}\xspace}

\def\pom {\ensuremath{\pm}\xspace}
\def\Chi {\ensuremath{\chi}\xspace}

\def\az    {\ensuremath{A_{0}}}

\def\thetakstar {\ensuremath{\theta_{\Kstar}}}

\def\cthetakstar{\ensuremath{\cos{\thetakstar}}}
\def\sthetakstar{\ensuremath{\sin{\thetakstar}}}

\newcommand{\cq}[1]{\cos^{2}{#1}}

\newcommand{\BABARPubYear}    {04}
\newcommand{\BABARPubNumber}  {46}

\newcommand{\SLACPubNumber} {10989}

\def\figurebox#1#2#3{%
    \def\arg{#3}%
    \ifx\arg\empty
    {\hfill\vbox{\hsize#2\hrule\hbox to #2{\vrule\hfill\vbox to #1{\hsize#2\vfill}\vrule}\hrule}\hfill}%
    \else
    {\hfill\epsfbox{#3}\hfill}%
    \fi}

\def\Chic  {\ensuremath{\chi_{c}}\xspace}
\def\Kstarone   {\ensuremath{K^*(892)}\xspace}
\def\chicJ  {\ensuremath{\chi_{cJ}}\xspace}

\begin{document}

 
 \par\vskip 0.8cm

\preprint{\babar-PUB-\BABARPubYear/\BABARPubNumber} 
\preprint{SLAC-PUB-\SLACPubNumber} 

\begin{flushleft}
\babar-PUB-\BABARPubYear/\BABARPubNumber\\
SLAC-PUB-\SLACPubNumber\\
\end{flushleft}

\title{
{\large \bf Search for Factorization-Suppressed \boldmath $\B\to\Chic\kaons$ \unboldmath Decays} 
}

%
\author{B.~Aubert}
\author{R.~Barate}
\author{D.~Boutigny}
\author{F.~Couderc}
\author{Y.~Karyotakis}
\author{J.~P.~Lees}
\author{V.~Poireau}
\author{V.~Tisserand}
\author{A.~Zghiche}
\affiliation{Laboratoire de Physique des Particules, F-74941 Annecy-le-Vieux, France }
\author{E.~Grauges-Pous}
\affiliation{Universitad Autonoma de Barcelona, E-08193 Bellaterra, Barcelona, Spain }
% ...
\collaboration{The \babar\ Collaboration}
\noaffiliation

\date{\today}

\begin{abstract}
We search for the factorization-suppressed decays 
 $\B\to\chiczero\kaons$ and 
 $\B\to\chictwo\kaons$, with \chiczero and \chictwo decaying into $\jpsi\g$, using 
 a sample of $124 \times 10^{6}$ \BB events collected with the \babar\
 detector at the \pep2\ storage ring of the Stanford Linear Accelerator
 Center. 
We find no significant signal and set upper bounds for the branching fractions.
\end{abstract}


\maketitle
Nonleptonic decays of heavy mesons are not easily described because
the process  involves quarks whose hadronization is 
not yet well understood.
The factorization hypothesis allows one to make some predictions~\cite{Bauer:1986bm}
 by assuming that a weak decay matrix element can be
described as the product of two independent hadronic currents.
Under the factorization hypothesis, $\B\to\ccbar K^{(*)}$ decays are
allowed when the \ccbar pair hadronizes to \jpsi, \psitwos or \chicone, but
suppressed when the \ccbar pair hadronizes to \chiczero or \chictwo~\cite{Suzuki:2002sq}.
Here,  $K^{(*)}$ represents either $K$ or \Kstar. 
In lowest-order Heavy Quark Effective Theory, there is no $J\ge2$
current to create the tensor \chictwo from the vacuum.
The decay rate to the scalar \chiczero is zero due to charge conjugation
invariance~\cite{hcGudrun}.

Belle has recently observed $\Bp\to\chiczero\Kp$ decays with a branching fraction (BF) 
of $(6.0^{+2.1}_{-1.8}\pm 1.1) \times 10^{-4}$~\cite{BelleChic0}
using  \chiczero decays to $\pip\pim$ or $\Kp\Km$.
\babar\ has confirmed the observation using the same decays with a branching
 fraction of $(2.7\pom 0.7)\times 10^{-4}$~\cite{Silvano}, somewhat lower than, but
 compatible with, the Belle measurement.
These results are of the same order of magnitude as the BF of the decay
$\Bp\to\chicone\Kp$  and are surprisingly large given the
expectation from factorization.
Using the hadronic \chiczero decays,
CLEO has obtained an upper limit on $\Bz\to\chiczero\Kz$ 
of  $5.0\times 10^{-4}$~\cite{CLEO2001}.
Non-factorizable contributions to $\Bp\to\chiczero\Kp$ decays due to
rescattering of intermediate charm states have been considered theoretically
\cite{colangelo}, and similar branching fractions are predicted for
decays to \chiczero and \chictwo.
No predictions are available for $B$ decays to $\Chi_{c(0,2)}\Kstar$,
but the branching fraction of decays to \Kstar may be expected to be
similar to the branching fraction of decays to $K$.
The measurement of $\B\to\Chi_{c(0,2)} K^{(*)}$ should improve our
 understanding of the limitations of factorization and of models that
 violate factorization.

In this Letter we report a search for the decays $\B\to\chicJ\kaons$, $J=0,2$, 
using the radiative decays $\chicJ\to\jpsi\g$, with branching
  fractions of $(1.18 \pm 0.14)\%$, $(20.2 \pm 1.7)\%$,
  respectively~\cite{ref:pdg}.
Since the radiative branching fraction for the \chiczero decay 
(including subsequent \jpsi decay to \ellell)
is much smaller than the
corresponding $\pip\pim$ or $\Kp\Km$ branching fractions, the search for the $\Bp\to\chiczero\Kp$ 
decay is less sensitive than previous searches, but it is free from the interference 
with the non-resonant decays to three mesons that affect the latter.
The data used in this analysis were obtained with the \babar\
detector at the \pep2\ storage ring, comprising an integrated
luminosity of 112\invfb of data taken at the \FourS resonance.

The \babar\ detector is described elsewhere~\cite{ref:babar}.
 Surrounding the interaction point, a five-layer double-sided silicon
vertex tracker (SVT) provides precise reconstruction of track angles
and \B-decay vertices. A 40-layer drift chamber (DCH) provides
measurements of the transverse momenta of charged particles.  An
internally reflecting ring-imaging Cherenkov detector (DIRC) is used
for particle identification (PID).  A CsI(Tl) crystal electromagnetic
calorimeter (EMC) detects photons and electrons.
The calorimeter is surrounded by a solenoidal magnet providing a 1.5-T field.
The flux return is instrumented with resistive plate chambers 
used for muon and neutral-hadron identification.

The channels considered here are $\B\to\Chic\kaons$ with 
$\Chic\to\jpsi\g$ and $\jpsi\to\ellell$, where $\ell$ is $e$ or $\mu$;
\kaon is \Kp or \KS($\to\pip\pim$);
 $\Kstarz\to\Kp\pim$ or $\KS\piz$;
 $\Kstarp\to\Kp\piz$ or $\KS\pip$; and $\piz\to\g\g$.
Charge-conjugate modes are included implicitly throughout this paper.
Event selection is optimized by maximizing $\epsilon/\sqrt{B}$, where
$\epsilon$ is the signal efficiency after all selection requirements
and $B$ the number of background events, estimated with $\FourS\to\BB$
and $\epem\to\qqbar$ Monte Carlo (MC) samples.

Candidate \jpsi mesons are reconstructed from a pair of oppositely
charged lepton candidates that form a good vertex.
Muon (electron) candidates are identified with a neural-network
(cut-based) selector and  loose selection criteria.
Electromagnetic depositions in the calorimeter in the polar-angle
range $0.410 < \theta_{lab} < 2.409\rad$ that are not associated with
charged tracks, have an energy larger than 30\mev, and a shower shape
consistent with a photon are taken as photon candidates.
For $\jpsi\to\epem$ decays, electron candidates are combined with nearby photon candidates 
in order to recover some of the energy lost through bremsstrahlung. 
The lepton-pair invariant mass must be in the range [2.95, 3.18] \gevcc for
both lepton flavors. 
The small remaining background is mainly due to \jpsi mesons not originating from \Chic decays.

We form \KS candidates from oppositely-charged tracks originating from
a common vertex with invariant mass in the range [487, 510] \mevcc.
The \KS flight length must be greater than 1\mm, and its direction in
the plane perpendicular to the beam line must be within 0.2\rad of the
\KS momentum vector.
Charged kaon candidates are identified with a likelihood selector,
based on information from the DIRC, and ${d E}/{d x}$ in the SVT and in the DCH.

A \piz candidate is formed from a pair of photon candidates with invariant mass in
the interval [117, 152] \mevcc and momentum greater than 350\mevc. 
\Kstar candidates are formed from $K\pi$ combinations with an
invariant mass in the range [0.85, 0.94] \gevcc.

The \jpsi, \KS, and \piz candidates are constrained to their
corresponding nominal masses~\cite{ref:pdg} to improve the resolution
of the measurement of the four-momentum of their parent \B-candidate.
The \Chic candidates are formed from \jpsi and photon candidates. The
photon is required to have an energy greater than 0.15\gev and not to be
part of \piz candidates in the mass range [0.125, 0.140]  \gevcc.

Candidate \B mesons are formed from \Chic and $K^{(*)}$ candidates.
Two kinematic variables are used to further remove incorrectly reconstructed \B candidates.
The first is the difference $\Delta E \equiv E^*_B - E^*_{beam}$ between
the \B-candidate energy and the beam energy in the \FourS rest
frame. In the absence of experimental effects, reconstructed signal
candidates have $\Delta E = 0$. 
The typical $\Delta E$ resolution is 20 \mev for channels with only charged tracks in the final state,
and  25 \mev, with a low   $\Delta E$ tail due to energy leakage in the calorimeter, for channels with a \piz.
The second variable is the beam-energy-substituted mass $\mes \equiv
(E^{*2}_{beam} -p^{*2}_B)^{1/2}$, where $p^{*}_B$ is the momentum of the
\B-candidate in the \FourS rest frame.
The energy substituted mass \mes should peak at the \B meson mass, 5.279\gevcc.
Typical resolution for $\Delta E$ is 2.7 \mevcc.
For the signal region, $\Delta E$ is required to be in the range
$[-35, +20] \mev$ for channels involving a \piz, and within $\pm 20 \mev$ otherwise.
We require \mes to be in the range [5.274, 5.284] \gevcc.
If more than one \B candidate is found in an event, the one having the
smallest $|\DeltaE|$ is retained.

The observation of \chictwo could be complicated by the presence of
the prominent \chicone peak. 
This is mitigated by measuring the spectrum in the variable
$m_{\ellell \g}-m_{\ellell}$.
The efficiencies obtained from fits to the mass difference distribution for 
exclusive MC samples, where one \B decays to the final state under
consideration and the other inclusively, are given in Table
\ref{tab:chic1:eff}. 
The \chictwo meson has a natural width of just 2 MeV~\cite{ref:pdg}
 and is therefore fitted with a Gaussian to account for detector
 resolution.
Since the \chiczero has a natural width of 10 MeV ~\cite{ref:pdg},
comparable to the mass resolution ($\sigma \approx 10 \mevcc$), we fit
the \chiczero peak with the convolution of Breit-Wigner and Gaussian
shapes.
\begin{table}[htbp]
\caption{Efficiencies from fits of exclusive MC distributions of $m_{\ellell \g}-m_{\ellell}$, 
with statistical uncertainty.
 \label{tab:chic1:eff}}
\begin{center}  
\begin{tabular}{cccc}
\hline
\hline
 & \chictwo & \chiczero \\
\hline
 \Kstarz ($\Kp \pim$) &       0.071 \pom     0.001 &   0.066 \pom     0.001 \\
 \Kstarz ($\KS \piz$) & 0.031 \pom     0.001 &      0.020 \pom     0.001  \\
 \KS &   0.158 \pom     0.001 &     0.126 \pom     0.001  \\
 \Kstarp ($\Kp \piz$) &    0.036 \pom     0.001 &  0.031 \pom     0.001 \\
 \Kstarp ($\KS \pip$) &    0.065 \pom     0.001 &     0.062 \pom     0.001    \\
 \Kp &   0.144 \pom     0.001 &      0.117 \pom     0.002 \\
\hline
\hline
\end{tabular}
\end{center}
\end{table}

\begin{figure}[htbp]
\begin{center}
\includegraphics[width=\linewidth]{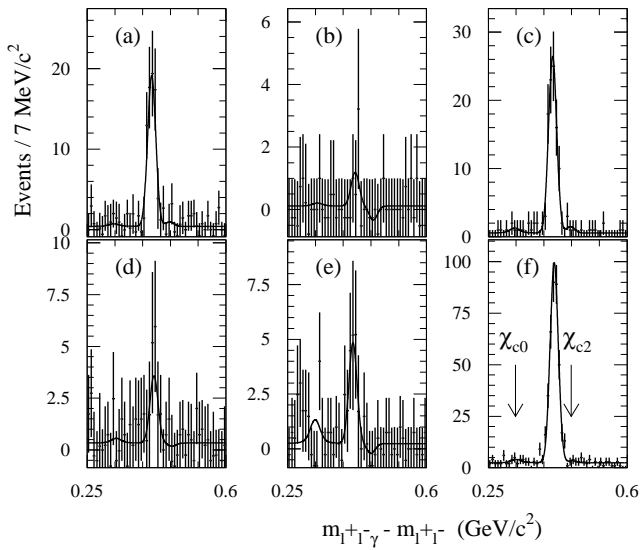}
\caption{Distribution of $m_{\ellell \g}-m_{\ellell}$ for data, with
 NR subtraction for  final states of the strange meson
(a) $\Kp \pim$,
(b)  $\KS \piz$,
(c)  \KS,
(d)  $\Kp \piz$,
(e)  $\KS \pip$,
(f)  \Kp.
The fit is described in the text.
The arrows on plot (f) show the expected positions of the \chiczero and \chictwo peaks.
\label{fig:data}}
\end{center}
\end{figure}

Studies of MC samples show that most of the background
events in the $\Chic\Kstar$ channels are due to non-resonant (NR) 
$\B\to\Chic(\jpsi\g)\kaon\pion$ decays.
After the NR events are removed from the MC background sample, the
expected background with a genuine $\Chic\to\jpsi\g$ decays is $0.2
\pom 0.2$ event for the $\chictwo\Kstarz(\Kp\pim)$ and
$\chictwo\Kstarp(\Kp\piz)$ modes, and $0.0\pom 0.2$ for all other
channels.
We correct for the presence of NR decays 
with the following procedure. 
The $m_{\ellell \g}-m_{\ellell}$ distribution for events in a nearby
sideband ($1.1 < m_{K \pi} < 1.3\gevcc$) is subtracted from the
distribution for events in the signal region 
($0.85 < m_{K \pi} < 0.94\gevcc$), after scaling the sideband
distribution by a factor $r = 0.26 \pom 0.04$.
The quantity $r$, obtained from MC simulation, is the ratio of NR
 events under the peak to the number in the sideband.
NR-subtracted distributions of $m_{\ellell \g}-m_{\ellell}$ are shown 
in Fig.\,\ref{fig:data}.
These plots show the presence of the factorization-allowed
\chicone but no significant signals for the factorization-suppressed \chiczero or \chictwo. 
No  \chiczero or \chictwo signal is observed in the sideband region.

\begin{table}[htbp]
\caption{Event yields with statistical uncertainties from the fits of Fig. 
\ref{fig:data}. 
 \label{tab:evts:number}}
\begin{center}  
\begin{tabular}{lccc}
\hline
\hline
 & \chictwo & \chiczero \\
\hline
 \Kstarz ($\Kp \pim$) &      2.0 \pom  1.6 &     1.7 \pom  2.1 \\
 \Kstarz ($\KS \piz$) &     -1.6 \pom  4.3 &     0.5 \pom  0.3 \\
 \KS &                       3.4 \pom  1.8 &     3.9 \pom  3.8 \\
 \Kstarp ($\Kp \piz$) &     -0.5 \pom  0.2 &     1.1 \pom  2.2 \\
 \Kstarp ($\KS \pip$) &     -1.9 \pom  1.2 &     5.9 \pom  3.7 \\
 \Kp &                       3.7 \pom  4.4 &     8.8 \pom  6.6 \\
\hline
\hline
\end{tabular}
\end{center}
\end{table}

The branching fractions are computed from 
 $BF = N_{S}/(N_{B}  \epsilon  f)$,
where $N_{S}$ is the number of signal events obtained from fitting the $m_{\ellell \g}-m_{\ellell}$
distribution (Table~\ref{tab:evts:number}),
 $N_B$ is the number of produced \BB events, 
$\epsilon$ is the selection 
efficiency (Table~\ref{tab:chic1:eff}) and 
$f$ is the product of secondary branching fractions of the \B daughters. 
The free parameters in the fits are the size of a constant background, the overall scale
of  $m_{\ellell \g}-m_{\ellell}$, and the amplitudes of the resonant peaks.
The fixed parameters are the \chiczero natural width, the
\chiczero--\chicone and \chictwo--\chicone mass differences 
($-95.4$ and +45.7\mevcc, respectively) all taken from Ref.~\cite{ref:pdg},
and the mass resolution.
The mass resolution, 10.2 \pom 0.4\mevcc, is measured with \chicone data
 and is 
assumed to be the same for the three \Chic states.
Performing 
 such fits to an inclusive $\FourS\to\BB$ MC sample, we verify that the NR events are
subtracted correctly, and that the proximity of the \chicone does not
induce any significant bias on the measurement of the nearby \chictwo.

Based on studies of \B\to\jpsi\Kstar decays~\cite{Aubert:2001pe},
the NR $K\pi$ component appears to be in an $S$-wave state, with
an unknown relative phase $\phi$ with respect to the main \Kstarone
$P$-wave peak.
As no signal is found, the systematic uncertainty due to the unknown relative
phase is estimated here with a MC-based method.
The $K-\pi$ invariant mass is fitted with an amplitude that is the sum
of a non-relativistic Breit-Wigner and an amplitude with a constant
phase and the square of which has a quadratic dependence on $m_{K\pi}$.
%
\begin{equation}
p(m_{K \pi}) = \left| \frac{a}{m_{\Kstar}-m_{K \pi}-i\Gamma/2}+b(m_{K
   \pi}) e^{i \phi}\right|^2\,,
\label{eq:BreitPlusNR}
\end{equation}
where $a$ and $b$ are real quantities and $m_{\Kstar}=892\mevcc$. The slow variation of the
phase of the $S$ wave with $m_{K \pi}$ is neglected here.
%
The free parameters in the fit are the three degrees of freedom of the
quadratic dependence of $b$, the magnitude of the signal, and the
relative phase $\phi$.
As the sideband is dominated by the NR contribution, no
attempt is made to subtract the few combinatorial events.
The fact that the phase $\phi$ is unknown is dealt with by randomly generating
samples of events distributed as above for each value of $\phi$, and
applying NR subtraction.  
The number of events $N(\phi)$
thus measured is normalized to that obtained with the phase value $\phi_0$ 
obtained in the fit. The ratio $R=N(\phi)/N(\phi_0)$ shows a sinusoidal dependance.
The average value is 1.44 with a deviation of $\pom 35\%$, giving an RMS
relative uncertainty of $\pom 20\%$, which we will
assume as systematic uncertainty (due to the interference with the
NR component).

In the case of decays to the tensor \chictwo, the efficiency depends
on the intensity fractions to each of three polarization states.
The efficiency is mainly sensitive to the value of the \Kstar helicity angle $\thetakstar$,
because small values of $\thetakstar$ occur for low momentum pions.
The selection efficiency therefore depends, to first order, on the
polarization of the \Kstar population, through the angular distribution:
\begin{equation}
\frac{1}{\Gamma} 
\frac{d \Gamma}{d  \cthetakstar}=
\frac{3}{4} \left[ (1-\cq{\thetakstar}) + \az (3\cq{\thetakstar}-1) \right],
\label{1Dcthetakstar}
\end{equation}
where \az\ is the fraction of longitudinal \Kstar polarization.
The average efficiency is
\begin{equation}
\langle
\varepsilon
\rangle =
\int 
\frac{1}{\Gamma} 
\frac{d \Gamma}{d  \cthetakstar}
\varepsilon (\thetakstar)
d  \cthetakstar 
 = a+ \az b,
\label{1DEffcthetakstar}
\end{equation}
where
$a=
\frac{3}{4}
\int 
 (1-\cq{\thetakstar})
\varepsilon (\thetakstar)\sthetakstar
d  \thetakstar,
$
and
$b=
\frac{3}{4}
\int (3\cq{\thetakstar}-1)
\varepsilon (\thetakstar)\sthetakstar
d  \thetakstar,
$
where $\varepsilon (\thetakstar)$ is obtained from MC.
The values of $a$ and $b$ are  shown in Table~\ref{tab:coefs:efficiency:angle}. 

When no signal is observed, as is the case here, the polarization is
 unknown.
We assume an unpolarized decay and we estimate the efficiency as 
$(a+ 0.5 b) \pm (|b|/\sqrt{12})$.
\begin{table}[htbp]
\caption{Coefficients for the calculation of amplitude-dependent
average efficiency for the $\chictwo \Kstar$ channels (\%).
\label{tab:coefs:efficiency:angle}}
\begin{center}    
\begin{tabular}{cccc}
\hline 
\hline
           &   $a$ &   $b$ &  Efficiency     \\ 
\hline 
 \Kstarz ($\Kp \pim$)  &      8.68 &  -1.40 &   7.98\pom 0.40  \\
 \Kstarz ($\KS \piz$) &      4.25 &  -1.66 &   3.43\pom 0.48  \\
 \Kstarp ($\Kp \piz$)  &      5.05 &  -1.79 &   4.16\pom 0.52  \\
 \Kstarp ($\KS \pip$) &     7.83 &  -1.84 &   6.92\pom 0.53  \\
\hline 
\hline
\end{tabular}
\end{center}
\end{table}
The branching fraction measurements
reported here are affected by the systematic uncertainties described in
what follows.
The relative uncertainty on the number of \BB events is 1.1\%.
The secondary branching fractions and their uncertainty are taken from
 Ref.~\cite{ref:pdg}.
Other estimated  uncertainties  are:
 tracking efficiency,  1.3\% per track added linearly;
\KS reconstruction,  2.5\%;
 selection of the \g from the $\chi_c$ decays, 2.5\%;
 \piz selection, 5.0\%;
 PID efficiency, 3.0\%.
For each mass peak and for $\Delta E$, the uncertainty of the central
 value and of the width of the peaks are measured with the \chicone
 channels.
These quantities are used to estimate the efficiency uncertainty from this source.
The ratio of \Bz to \Bp production in \FourS decays is assumed to be unity.
The related uncertainty is small~\cite{Aubert:2004ur}
and is neglected here.
A summary of the multiplicative contributions to the systematics can
be found in Table\,\ref{tab:SumSyst}.
\begin{table}[!t]
\caption{Summary of the multiplicative 
systematic uncertainties in percent.
The first eight rows are in common to decays to \chiczero and \chictwo.
\label{tab:SumSyst}}
\begin{center}    
\begin{tabular}{lcccccc}
\hline 
\hline
& $\Kp \pim$ &  $\KS \piz$ &   $\Kp \piz$ &  $\KS \pip$ &  \Kp &   \KS \\
\hline 
Number of \B's & 1.1 & 1.1 & 1.1 & 1.1 & 1.1 & 1.1 \\
Tracking   & 5.2 & 2.6 & 3.9 & 3.9 & 3.9 & 2.6 \\
\KS        & --  & 2.5 & -- & 2.5 & -- & 2.5 \\
Neutrals   & 2.5 & 7.5 & 7.5 & 2.5 & 2.5 & 2.5 \\
PID        & 3.0 & 3.0 & 3.0 & 3.0 & 3.0 & 3.0 \\
Sample selection
           & 7.7 & 13.1 & 11.6 & 8.2 & 6.5 & 6.3 \\
MC statistics &
 1.4 & 2.9 & 1.7 & 1.8 & 1.3 & 1.3 \\
S-wave Phase &  20.0 &  20.0 &  20.0 &  20.0 & -- &  -- \\
\hline 
\chiczero  second. BF & 11.9   & 11.9  & 11.9  & 11.9  & 11.9  & 11.9 \\
\hline 
Total for \chiczero & 
      25.4   &   28.3   &   27.6  &    25.5    &  14.8   &   14.6 \\
\hline 
\chictwo second. BF & 8.5  & 8.5  & 8.5  & 8.5  & 8.5 & 8.5 \\ 
Polarization & 5.1 & 14.0 & 12.4 & 7.7 & -- & -- \\
\hline 
Total for \chictwo &  
      24.5   &   30.5  &    29.1  &    25.3   &   12.2  &    12.0 \\
 \hline 
\hline
\end{tabular}
\end{center}
\end{table}
In addition to these multiplicative contributions there is a small
contribution from the uncertainty on $r$ for the NR background subtraction.

Combining the measurements of the \Kstar sub-modes, and with the  approximation
that the multiplicative efficiencies for each \Kstar sub-mode are
fully correlated, we obtain the branching fractions for the factorization-suppressed modes listed in Table 
\ref{tab:bf:meas}.
As a cross check,
the results for the allowed \chicone are found to be compatible with 
 those of a recent  analysis~\cite{Philippe}
  optimized for that decay.
We obtain upper bounds on the BFs at 90\% confidence level (C.L.)
assuming Gaussian statistics for the statistical uncertainties and
taking into account the systematic uncertainties.
We have used a Bayesian method with uniform prior for positive BF values
in the derivation of these limits.
The upper limits obtained for decays to \chiczero are larger than for  \chictwo due to
the smaller \chiczero  radiative BF. For  $\Bp\to\chiczero \Kp$ they are
compatible with the previous measurements
\cite{BelleChic0,Silvano}.

$\B\to\Chi_{c(0,2)}\kaons$ production requires non-factorizable
 contributions. $\Bp\to\chiczero\Kp$ decays have been previously
 observed.
Colangelo {\em et al.}~\cite{colangelo}
explain this with rescattering effects and
predict a similar rate for  $\B\to\chictwo K$.
This is not observed.
The upper limits obtained for decays to \chictwo are approximately one
order of magnitude lower than the branching fractions of the
observed $\Bp\to\chiczero\Kp$ decays.
Furthermore, we find no evidence for the decays  $\B\to\Chi_{c(0,2)}\Kstar$.
\begin{table}[!h]
\caption{Upper limits  at 90\% C.L. and 
measured branching fractions  (in pararentheses) in units of $10^{-4}$.\label{tab:bf:meas}
}
\begin{center}
\begin{tabular}{ccccc}
\hline 
\hline
 & \chictwo & & \chiczero & \\ 
\hline 
\Kstarz & 
  0.36 &   (0.14 \pom  0.11 \pom  0.14) & 7.7 &  (3.8 \pom  2.6 \pom 1.5) \\
\Kstarp & 
  0.12 &    (-0.15 \pom  0.05 \pom  0.14) & 28.6 &  (13.5 \pom  9.6 \pom 5.3) \\
\Kp &
  0.30 &    (0.09 \pom  0.10 \pom  0.11) & 8.9 &  (4.4 \pom  3.3 \pom 0.7) \\
\Kz &
  0.41 &     (0.21 \pom  0.11 \pom  0.13) & 12.4 &  (5.3 \pom  5.0 \pom 0.8) \\
\hline 
\hline
\end{tabular}
\end{center}
\end{table}

We are grateful for the excellent luminosity and machine conditions
provided by our \pep2\ colleagues, 
and for the substantial dedicated effort from
the computing organizations that support \babar.
The collaborating institutions wish to thank 
SLAC for its support and kind hospitality. 
This work is supported by
DOE
and NSF (USA),
NSERC (Canada),
IHEP (China),
CEA and
CNRS-IN2P3
(France),
BMBF and DFG
(Germany),
INFN (Italy),
FOM (The Netherlands),
NFR (Norway),
MIST (Russia), and
PPARC (United Kingdom). 
Individuals have received support from CONACyT (Mexico), A.~P.~Sloan Foundation, 
Research Corporation,
and Alexander von Humboldt Foundation.


\end{document}